\documentclass[twocolumn,pra,aps,showpacs,superscriptaddress,floatfix,showkeys,nofootinbib,10pt]{revtex4-2}

\usepackage{amsmath,amsfonts,amssymb,amsthm}
\usepackage{graphicx}
\usepackage{color}
\usepackage{xcolor}

\newcommand{\ket}[1]{ | \, #1 \rangle}

\newcommand{\be}{\begin{equation}} \newcommand{\ee}{\end{equation}}
\newcommand{\ba}{\begin{aligned}} \newcommand{\ea}{\end{aligned}}

\DeclareMathOperator{\Tr}{Tr}

\DeclareRobustCommand\openone{\leavevmode\hbox{\small1\normalsize\kern-.33em1}}%

\newcommand\bigforall{\mbox{\Large $\mathsurround=0pt\forall$}}

\begin{document}

\title{Experimental certification of more than one bit of quantum randomness in the two inputs and two outputs scenario}

\author{Alban Jean-Marie Seguinard}
\affiliation{Department of Physics, Stockholm University, S-10691 Stockholm, Sweden}

\author{Amélie Piveteau}
\affiliation{Department of Physics, Stockholm University, S-10691 Stockholm, Sweden}

\author{Piotr~Mironowicz} \email{piotr.mironowicz@gmail.com}
\affiliation{Department of Physics, Stockholm University, S-10691 Stockholm, Sweden}
\affiliation{Department of Algorithms and System Modeling, Faculty of Electronics, Telecommunications and Informatics, Gda\'nsk University of Technology, Poland}
\affiliation{International Centre for Theory of Quantum Technologies, University of Gda\'{n}sk, Wita Stwosza 63, 80-308 Gda\'{n}sk, Poland}

\author{Mohamed Bourennane}
\affiliation{Department of Physics, Stockholm University, S-10691 Stockholm, Sweden}

\date{\today}

\begin{abstract}
One of the striking properties of quantum mechanics is the occurrence of the Bell-type non-locality. They are a fundamental feature of the theory that allows two parties that share an entangled quantum system to observe correlations stronger than possible in classical physics. In addition to their theoretical significance, non-local correlations have practical applications, such as device-independent randomness generation, providing private unpredictable numbers even when they are obtained using devices derived by an untrusted vendor. Thus, determining the quantity of certifiable randomness that can be produced using a specific set of non-local correlations is of significant interest. In this paper, we present an experimental realization of recent Bell-type operators designed to provide private random numbers that are secure against adversaries with quantum resources. We use semi-definite programming to provide lower bounds on the generated randomness in terms of both min-entropy and von Neumann entropy in a device-independent scenario. We compare experimental setups providing Bell violations close to the Tsirelson's bound with lower rates of events, with setups having slightly worse levels of violation but higher event rates. Our results demonstrate the first experiment that certifies close to two bits of randomness from binary measurements of two parties.
\end{abstract}


\maketitle

\section{Introduction}

Randomness is one of the basic resources in information processing. Randomly generated numbers find applications in areas such as cryptography, where they are one of the key elements of protocols such as Data Encryption Standard (DES) or Advanced Encryption Standard (AES). The standard document \textit{RFC 4086 - Randomness Requirements for Security}~\cite{rfc4086} lists such fields of application as creating private keys for algorithms used in digital signatures, keys and initialization values for encryption, generating secure PINs and passwords, keys for MAC (Message Authentication Code) algorithms or nonces, \textit{i.e.} numbers that are being used just once in a cryptographic communication, and cannot be further reused.

It is a known fact that using classical computers, which operate on deterministic algorithms, it is not possible to generate truly random numbers, but only sequences of pseudo-random values, which at first glance resemble truly random numbers, but are not able to guarantee to be unpredictable. Anyone who knows the algorithm used to create them and its input parameters, \textit{i.e.} the so-called seed, can determine all the numbers that will ever be obtained using the deterministic generator.

The situation is conceptually different in quantum mechanics since its essence is processes that behave in a non-deterministic way. Thus, many quantum phenomena have intrinsic randomness. The so-called coherence of quantum states can be shown to be directly related to the complete unpredictability of certain quantities~\cite{yuan2015intrinsic}.

Nevertheless, verifying that a quantum device works as expected is a much more difficult issue than checking the correctness of deterministic algorithms. Typically, we cannot tell if a quantum device behaves exactly as designed, and imperfections in both quantum states and measurements can cause the entire process to lose quantum characteristics, such as Bell inequality violation~\cite{bell1964einstein}. The situation is even worse due to the complexity of quantum devices and their finesse, as we are often unable to even check whether the components we use have not been intentionally manipulated by a malevolent adversary.

An important milestone towards solving this problem was the emergence of the so-called device-independent approach~\cite{mayers1998quantum}, which allows the assessment of the fidelity of a quantum device based on its visible external behavior. The early works containing experimental implementations of quantum randomness protocols presented a proof of concept~\cite{pironio2010random}, but they were not very efficient in terms of the generation ratio. Recently, a new theorem called Entropy Accumulation Theorem (EAT) was introduced and proven~\cite{dupuis2020entropy,arnon2019simple,arnon2018practical}. This theorem allowed for a determination of the amount of randomness which is certified to be generated by a particular quantum device.

One of the important problems is how to obtain the largest possible amount of certified randomness in terms of individual rounds by using a given device with the simplest configuration of settings and outcomes. It is easy to see that a device involving two components, $A$ and $B$, each generating one output bit, allows for a maximum of 2 random bits generation per round. The first protocols allowing a certification of the maximal amount of two bits where one of the parties use three measurement settings are presented in~\cite{mironowicz2013robustness}.

Recently the simplest known protocol for certification of two bits, involving only two binary measurements of two parties have been introduced~\cite{wooltorton2022tight}. In this work, we will present its experimental implementation, along with the analysis of certified randomness using various numerical techniques to provide a lower bound on the randomness generated per round~\cite{brown2019framework,brown2019constructions}.

\section{Methods}

For a given behavior of a quantum device, our task is to specify lower bounds on the generated randomness. This task is called randomness certification. To this end, it is necessary to perform a complex optimization taking into account all devices implementing this behavior allowed by the laws of quantum physics. This optimization is essentially a consideration of the set of all possible probability distributions obtainable by quantum devices. The are no known tools to optimize accurately over this set of probability distributions. Fortunately, there are approximate techniques that determine the so-called relaxations of the set of all possible distributions of quantum probabilities. It turns out that if the optimization over probability distributions is allowed to cover a set slightly wider than that allowed by quantum mechanics, the optimization problem can be dealt with efficiently using convex optimization techniques, in particular, semi-definite programming (SDP)~\cite{vandenberghe1996semidefinite}. One of the most widely used methods is the so-called Navascués-Pironio-Acin (NPA)~\cite{navascues2007bounding,navascues2008convergent}.

The NPA is used to optimize the probability of guessing the value of a random number by an adversary~\cite{impagliazzo1989pseudo}. The guessing probability is directly related to the so-called min-entropy~\cite{konig2009operational}. It can be shown that the min-entropy is a lower bound on the Shannon entropy of a given random variable. Thus, the values obtained using the NPA are suitable for certification of the generated randomness from quantum devices. The NPA is limited only to optimize functions that are linear expressions of probabilities.

Let us consider the set of all probability distributions $\{P(a,b|x,y)\}$, where $a$ and $b$ are the outcomes of the measurements performed by Alice and Bob, when their measurement settings are $x$ and $y$, respectively. One of the proposed protocols from~\cite{mironowicz2013robustness} used the following Bell operator as a randomness privacy certificate:
\begin{equation}
	\label{eq:modCHSH}
	C(0,1) + C(0,2) + C(1,0) + C(1,1) - C(1,2),
\end{equation}
where the correlators are defined as follows:
\begin{equation}
	\begin{aligned}
		C(x,y) &\equiv P(0,0|x,y) + P(1,1|x,y) \\
		&- P(0,1|x,y) - P(1,0|x,y).
	\end{aligned}
\end{equation}
One may note that~\eqref{eq:modCHSH} consists of a well-known CHSH expression~\cite{clauser1969proposed} plus an additional term. The maximal value allowed in quantum mechanics, \textit{i.e.} the Tsirelson bound, of~\eqref{eq:modCHSH} is $1 + 2 \sqrt{2}$. When the Tsirelson bound is achieved, then the quantum state and all measurement operators are uniquely determined~\cite{vsupic2020self}, and for the pair of settings $x=0$, $y=0$, the measurement results are uniformly distributed, $\bigforall{a,b} P(a,b|0,0) = 0.25$. Protocols that employ a particular setting for randomness generation are called spot-checking protocols~\cite{miller2017universal}. A full analysis of the protocol of randomness generation using~\eqref{eq:modCHSH} including randomness accumulation and extraction aspect has been presented in~\cite{brown2019framework}.

Initially, SDP was used to produce an approximation of the matrix logarithm function~\cite{fawzi2019semidefinite}, resulting in an SDP for efficient optimization of expressions on the so-called quantum relative entropy~\cite{fawzi2018efficient}. Furthermore, this method have be used to determine the lower bounds of the conditional von~Neumann entropy certified in the device-independent approach~\cite{brown2021device} using the extended NPA~\cite{pironio2010convergent} with the NCPOL2SDPA tool~\cite{wittek2015algorithm}.

The technique can be applied to calculate a lower bound on the conditional von~Neumann entropy
\begin{equation}
	\label{eq:Habxye}
	H(a,b|x=x^{*},y=y^{*},e),
\end{equation}
where $e$ represents any sort of knowledge (classical or quantum) that an adversary may possess if governed by the laws of quantum mechanics. Let $Q_A$, $Q_B$, and $Q_E$ be the Hilbert spaces of devices of Alice, Bob, and adversary, respectively, and $\rho_{Q_A, Q_B, Q_E}$ their shared tri-partite quantum system. Let $\{ \{ M_{a|x} \}_a \}_x$ and $\{ \{ N_{b|y} \}_b \}_y$ denote operators of the positive operator valued measurements (POVMs) performed by Alice and Bob, respectively. This employs the Gauss-Radau quadrature rule to lower bound~\eqref{eq:Habxye}. Let $w_i$ and $t_i$ be the nodes and weights defined by this quadrature. A lower bound can be obtained from~\cite{Brown2022}:
\begin{equation}
	\label{eq:opt}
	\sum_{i} c_i \sum_{a,b=0,1} \inf_{\substack{Z_{a,b} \in B(Q_E), \\ \text{cond}(P)}} \left( 1 + F[M_{a|x^{*}}, N_{b|y^{*}}, Z_{a,b}, t_i] \right),
\end{equation}
where $F[M_{a|x^{*}}, N_{b|y^{*}}, Z_{a,b}, t_i]$ is defined as $\Tr \left[ \rho_{Q_A, Q_B, Q_E} \left( O_1 +  O_2 \right) \right]$.
In~\eqref{eq:opt} $\text{cond}(P)$ expresses that the probability distribution $P(a,b|x,y) \equiv \Tr[\rho_{Q_A,Q_B} M_{a|x^{*}} \otimes N_{b|y^{*}}]$ satisfies a certain, specified by the protocol, set of linear constraints. The operators $O_1$ and $O_2$ are defined by
\begin{subequations}
	\begin{equation}
		O_1 \equiv M_{a|x^{*}} \otimes N_{b|y^{*}} \otimes \left( Z_{a,b} + Z_{a,b}^{\dagger} + (1 - t_i) Z_{a,b} Z_{a,b}^{\dagger} \right),
	\end{equation}
	\begin{equation}
		O_2 \equiv t_i \left( \openone_{Q_A Q_B} \otimes Z_{a,b} Z_{a,b}^{\dagger} \right).
	\end{equation}
\end{subequations}
$c_i$ are coefficients calculated from the Gauss-Radau quadrature as $c_i \equiv w_i / (t_i \log(2))$.
The index $i$ in the summation~\eqref{eq:opt} takes the values indexing the nodes in the quadrature, omitting the last one.

To certify the randomness, we employed the recently announced two families Bell expressions~\cite{wooltorton2022tight}. First of them is, parametrized by $\delta \in (0, \pi/6]$, defined as
\begin{equation}
	\label{eq:Idelta}
	I_\delta \equiv C(0,0) + \frac{1}{\sin{\delta}} \left( C(0,1) + C(1,0) \right) - \frac{1}{\cos{2 \delta}} C(1,1).
\end{equation}
The members of this family have self-testing properties, use two settings for each party, and can certify two bits of randomness for the measurement settings $x^{*} = y^{*} = 0$.

The second family, parametrized by $\gamma \in [0, \pi/12]$ defines the Bell expressions:
\begin{equation}
	\begin{aligned}
		\label{eq:Jgamma}
		J_\gamma &\equiv C(0,0) + \\
		&\left( 4 \cos^2{\left[ \gamma + \pi / 6 \right]} -1 \right) \left( C(0,1) + C(1,0) - C(1,1) \right).
	\end{aligned}
\end{equation}
These Bell expressions also use two settings and have self-testing properties, yet in most cases do not certify two bits of randomness.

The Tsirelson bounds for~\eqref{eq:Idelta} and~\eqref{eq:Jgamma} are $I_\delta^Q \equiv 2 \cos^3{\delta} / \left( \cos{(2 \delta)} \sin{\delta} \right)$, and $J_\gamma^Q \equiv 8 \cos^3{[\gamma + \pi / 6]}$, respectively.
The relative Bell value is defined as $I_\delta^{exp} / I_\delta^Q$ and $J_\gamma^{exp} / J_\gamma^Q$, where $I_\delta^{exp}$ and $J_\gamma^{exp}$ are the values of the Bell expressions~\eqref{eq:Idelta} and~\eqref{eq:Jgamma} obtained in the experiment, respectively. The relative Bell value attains the value of $1$ in the noiseless cases. For correlation-based Bell expressions, like those analyzed in this paper, if $\eta$ is the relative value of the Bell expression, and the relative value $\eta$ is attained with the noised state:
\begin{equation}
	\eta \rho_{Q_A,Q_B}^{optimal} + (1-\eta) \rho_{Q_A,Q_B}^{white},
\end{equation}
where $\rho_{Q_A,Q_B}^{optimal}$ and $\rho_{Q_A,Q_B}^{white}$ are the quantum state providing the Tsirelson bound and the maximally mixed state, respectively.

\section{Results}

In this section, we describe the experimental setup, and report regarding the analysis of the randomness generated in the series of experiments.

\subsection{Experimental setup}

\begin{figure}
	\includegraphics[width=\columnwidth]{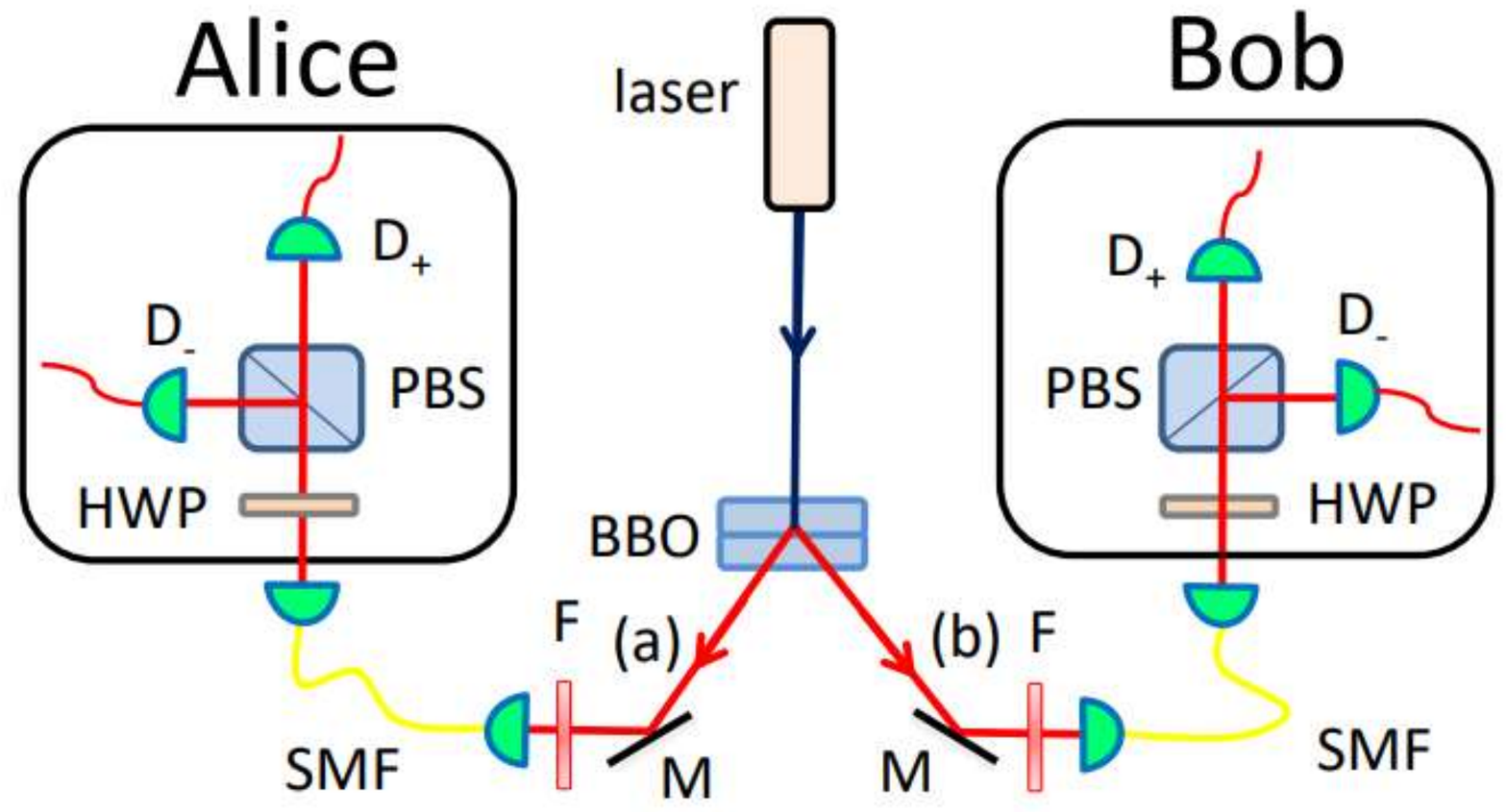}
	\caption{Experimental setup. Entangled photons pairs are generated through SPDC process. The signal is filtered. The two stations aof measurements are composed each by an half-wave-plate (HWP) and a polarization beam spliter (PBS). (See main text for details)}
	\label{fig:setup}
\end{figure}

Ultraviolet light centered at a wavelength of $390$~nm is focused onto two $2$~mm thick $\beta$ barium borate (BBO) nonlinear crystals placed in interferometric configuration to produce photon pairs emitted into two spatial modes (a) and (b) through the second order degenerate type-I spontaneous parametric down-conversion process. The spatial, spectral, and temporal distinguishability between the down-converted photons is carefully removed by coupling to single-mode fiber, passed through narrow-bandwidth interference filters (F) and quartz wedges respectively. We have realized these quantum protocols by using polarization entangled pairs of photons $\ket{\phi+}  = \ket{HH} + \ket{VV}$.

The measurements for Alice are performed by a half-wave plate (HWP) oriented $\theta_{A0}$ or $\theta_{A1}$, and the measurements for Bob are performed by an HWP-oriented $\theta_{B0}$ or $\theta_{B1}$. The polarization measurement was performed using PBS and single-photon detectors (D) placed at the two output modes of the PBS. Our detectors are actively quenched Si-avalanche photodiodes. All single-detection events were registered using a VHDL-programmed multichannel coincidence logic unit, with a time coincidence window of $1.7$ ns.

We performed the experiment for the Bell expressions~\eqref{eq:Idelta} at a low rate (approximately $675$ two-photon coincidences per second). At these low rates, the multi-photon pairs emission are small and accidental events can be neglected. We benchmark the state preparation by measuring the average visibility in the diagonal polarization basis of $99.07$. Each of the measurement runs was taken 180 times with each run with a collection time of 250 seconds. We have also performed a state tomography to estimate the fidelity of the state and obtained $99.63 \pm 0.04$.

For the Bell expressions~\eqref{eq:Jgamma}, the rate was around 7$80$ two-photon coincidences per second with 160 measurements of 250 seconds and the visibility in the diagonal polarization basis of $99.13\%$. The fidelity of the state obtained by state tomography is $99.75 \pm 0.02$. For these two experiments, the total average number of events is around 120 000 000.

We have also performed an experiment for the Bell expressions~\eqref{eq:Idelta} at a higher rate of 2000 two-photon coincidences per second, the visibility was on average $98.73$ in diagonal polarization basis. Each of the measurement runs was taken 100 times with each run with a collection time of 250 seconds.

To reduce experimental errors in the measurements, we used computer-controlled high-precision motorized rotation stages to set the orientation of wave-plates with repeatability precision $0.02^{\circ}$. The error was estimated for each of the experiments by taking the standard deviation of the measurements.

The experimental setup is illustrated in Fig.~\ref{fig:setup}. The angles of the HWPs for the experiments are provided in the Appenidx~\ref{app:Angles}.

\subsection{Certified randomness}

To calculate the guessing probability, we reflected the experimental results by imposing on the maximization a constraint that the value of the Bell expression is equal to the one observed in the experiment.

To calculate the conditional von~Neumann entropy, we have considered two different sets of constraints for the certification of the von~Neumann entropy using the optimization~\eqref{eq:opt}.
The standard approach~\cite{pironio2010random} is to impose a constraint that the value of the relevant Bell expression is equal to the one from the experiment.
A more involved method~\cite{mironowicz2013robustness,nieto2014using,bancal2014more,nieto2018device} is to constrain the optimization with more than one parameter. The purpose of this is to increase the amount of the certified randomness, at a price of more demanding error analysis for finite data sets, and complicated numerical calculations.
We imposed a constraint that each of the correlators $C(0,0)$, $C(0,1)$, $C(1,0$), and $C(1,1)$ are equal to those from the experiment.
Note that the latter constraints are stronger than the former one, as the Bell expressions~\eqref{eq:Idelta} and~\eqref{eq:Jgamma} are functions of correlators.

To be more precise, we have further relaxed the above constraints. 
We formulated the single parameter constraint in a form that the value of the relevant Bell expression is not smaller than the one from the experiment.
The constraints for more parameters we formulated in a manner that each of the correlators $C(0,0)$, $C(0,1)$ and $C(1,0)$ are not smaller than the one obtained in the experiment, and $C(1,1)$ is not greater than the one from the experiment.
It is easy to see that the minimization of the conditional von~Neumann entropy with equalities as a constraint will be lower bound by the minimization with inequalities. The reason behind this relaxation is that this improves the stability of the numerical optimization, as the feasible region has a wider interior than with equality constraints. Similarly, for the guessing probability calculations, we relaxed the equality with an inequality imposing a constraint of the optimization that the value of the Bell is not smaller than the one obtained in the experiment.

As mentioned, the method~\cite{brown2021device} requires specifying the number of nodes in the quadrature. We calculated both variants of constraints with $6$ nodes, and the optimization with the correlation constraints also with $8$ nodes. To improve the certification of entropy, one can increase the number of nodes, but this comes with the price of a longer optimization time.

\subsubsection{$I_\delta$ Bell expressions and high relative violation}

Firstly, we performed the experiment for the Bell expressions~\eqref{eq:Idelta}. We concentrated on the quality of the source, at the cost of the generated events rate. The experiment has been performed for $\delta = 0.45, 0.5, 0.52$. The obtained relative Bell values were $0.994$, $0.994$, and $0.997$, respectively, and the certified randomness is shown in Tab.~\ref{tab:prop1b}.

\begin{table}[hptb]
	{\scriptsize
		\begin{tabular}{|l|lll|l|}
			\hline
			& \multicolumn{3}{c}{von Neumann entropy} &      \\
			$\delta$ & Cor. (8 Radau) & Cor. (6 Radau) & Bell viol. (6 Radau) & $H_\infty$ \\ \hline
			$0.52$     & $1.88$        & $1.87$        & $1.77$        & $1.50$ \\
			$0.5$      & $1.77$        & $1.76$        & $1.64$        & $1.33$ \\
			$0.45$     & $1.77$        & $1.76$        & $1.61$        & $1.28$ \\ \hline
		\end{tabular}
	}
	\caption{Randomness certified by Bell expressions~\eqref{eq:Idelta} for the experiment concentrated on high relative violation of the Bell inequality.}
	\label{tab:prop1b}
\end{table}

We observed $675$ events per second, and thus the randomness generation rate for $\delta = 0.52$ is $1270$ bits of von~Neumann entropy or $1012$ bits of min-entropy, per second.

If only finite statistics are taken into account, one should consider also the uncertainty in evaluation \textit{e.g.} the Bell expression value. In the case of the performed experiment, the values are shown in Tab.~\ref{tab:violation_prop1b} with theoretical boundaries for comparison, for $\delta = 0.45, 0.5, 0.52$, respectively. The Gauss-Radau approximation with six nodes showed that this Bell violation allows certifying $1.54$, $1.58$, and $1.72$ bits of von~Neumann entropy, respectively, thus slightly less than the asymptotic case of Tab.~\ref{tab:prop1b}.
\begin{table}[hptb]
	\begin{tabular}{|l|lll|}
		\hline
		$\delta$ & $I_{\delta}^{Classique}$ & $I_{\delta}^{Quantique}$ & $I_{\delta}^{Experimental}$  \\ \hline
		$0.52$      &     $5$    &    $5.2$     & $5.179 \pm 0.006$      \\
		$0.5$      & $5.022$       & $5.218$      & $5.187 \pm 0.006$        \\
		$0.45$      &    $5.207$     &    $5,4$     & $5.366 \pm 0.007$         \\ \hline
	\end{tabular}
	\caption{The experimental values of the Bell expression~\eqref{eq:Idelta} for the experiment concentrated on high relative violation of the Bell inequality.}
	\label{tab:violation_prop1b}
\end{table}

\subsubsection{$J_\gamma$ Bell expressions}

Secondly, we investigated the Bell expressions~\eqref{eq:Jgamma}. We considered the value $\gamma = 0, \pi / 24, \pi / 12$. The obtained relative Bell values are $0.996$, $0.993$, and $0.994$, respectively. The certified randomness is shown in Tab.~\ref{tab:prop2}.

\begin{table}[hptb]
	{\scriptsize
	\begin{tabular}{|l|lll|l|}
		\hline
		& \multicolumn{3}{c}{von Neumann entropy} &      \\
		$\gamma$ & Cor. (8 Radau) & Cor. (6 Radau) & Bell viol. (6 Radau) & $H_\infty$ \\ \hline
		$0$            & $1.81$        & $1.80$        & $1.72$        & $1.43$ \\
		$\frac{\pi}{24}$     & $1.76$        & $1.75$        & $1.68$        & $1.21$ \\
		$\frac{\pi}{12}$     & $1.55$        & $1.54$        & $1.39$        & $0.98$ \\ \hline
	\end{tabular}
	}
	\caption{Randomness certified in the experiment by Bell expressions~\eqref{eq:Jgamma}.}
	\label{tab:prop2}
\end{table}

The observed event rate was $780$ per second, giving the randomness generation rate $1300$ bits of von~Neumann entropy or $1030$ bits of min-entropy, per second.

The violation of Bell's inequality is given in Tab.~\ref{tab:violation_prop2}.

\begin{table}[hptb]
	\begin{tabular}{|l|lll|}
		\hline
		$\gamma$ & $J_{\gamma}^{Classique}$ & $J_{\gamma}^{Quantique}$ & $J_{\gamma}^{Experimental}$  \\ \hline
		$0$      & $5$        & $5.19$        & $5.174\pm 0.007$         \\
		$\frac{\pi}{24}$      &   $3.55$     &    $3.99$   & $3.968 \pm 0.005$        \\
		$\frac{\pi}{12}$     &   $2$     &  $2.83$    & $2.811 \pm 0.003$         \\ \hline
	\end{tabular}
	\caption{The experimental values of the Bell expression~\eqref{eq:Jgamma}.}
	\label{tab:violation_prop2}
\end{table}

\subsubsection{$I_\delta$ Bell expressions and high event rate}

The third performed experiment concerned also the Bell expressions~\eqref{eq:Idelta}. We performed it for values $\delta = 0.5, 0.4, 0.3$ observing the relative Bell values $0.987$, $0.991$, and $0.991$, respectively. We show the certified randomness in Tab.~\ref{tab:prop1a}.

\begin{table}[hptb]
	{\scriptsize
		\begin{tabular}{|l|lll|l|}
			\hline
			& \multicolumn{3}{c}{von Neumann entropy} &      \\
			$\delta$ & Cor. (8 Radau) & Cor. (6 Radau) & Bell viol. (6 Radau) & $H_\infty$ \\ \hline
			$0.5$      & $1.50$        & $1.50$        & $1.26$        & $0.89$ \\
			$0.4$      & $1.59$        & $1.58$        & $1.41$        & $1.06$ \\
			$0.3$      & $1.52$        & $1.51$        & $1.21$        & $0.85$ \\ \hline
		\end{tabular}
	}
	\caption{Randomness certified by Bell expressions~\eqref{eq:Idelta} for the experiment concentrated on the high rate of the observed events.}
	\label{tab:prop1a}
\end{table}

The rate of observed events was $2000$ per second, so the randomness generation rate, when taking $\delta = 0.4$ is $3180$ bits of von~Neumann entropy or $2120$ bits of min-entropy, per second.

The violation of Bell's inequality is given in Tab.~\ref{tab:violation_prop1a}.

\begin{table}[h!]
	\begin{tabular}{|l|lll|}
		\hline
		$\delta$ & $I_{\delta}^{Classique}$ & $I_{\delta}^{Quantique}$ & $I_{\delta}^{Experimental}$  \\ \hline
		$0.5$      & $5.02$        & $5.22$        & $5.15 \pm 0,01$         \\
		$0.4$      & $5.57$       & $5.76$       & $5.71 \pm 0,01$        \\
		$0.3$      & $6.98$        & $7.15$        & $7.09 \pm 0.01$         \\ \hline
	\end{tabular}
	\caption{The experimental values of the Bell expression~\eqref{eq:Idelta} for the experiment concentrated on the high rate of the observed events.}
	\label{tab:violation_prop1a}
\end{table}

\section{Discusion and conclusions}

We have presented an experimental setup aiming to generate close to the maximum amount of randomness possible in the binary measurement setup with two parties. We have realized experiments for two different families of Bell expressions and obtained up to $1.88$ bits per round, that is close to the teoretical maximum of two bits.
We have also performed a comparison of different approaches to randomness, the von~Neumann and min-entropy. The min-entropy is smaller than the von~Neumann entropy, whereas some applications take advantage of the latter one.
Finally, we have shown, that it may be beneficial for the randomness generation rate, to increase the events rate at the cost of decreasing the quality of the quantum realization.
We find it interesting for a future research to answer the question of how does it influence the net gain of the randomness extraction? We expect that having close to two bits per elementary event will simplify the randomness extraction procedure, in terms of both requirements for the extractor's seed, and the extraction processing time. We leave the problem of employing the setup to full quantum randomness extraction for future work.

\section*{Acknowledgements}

This work was supported by the Knut and Alice Wallenberg Foundation through the Wallenberg Centre for Quantum Technology (WACQT), the Swedish research council (VR), and NCBiR QUANTERA/2/2020 (www.quantera.eu) an ERA-Net cofund in Quantum Technologies under the project eDICT.
The numerical calculation we conducted using NCPOL2SDPA~\cite{wittek2015algorithm}, and MOSEK Solver~\cite{mosek21}.

\appendix

\section{Angles}
\label{app:Angles}
Experiments 1 and 3 are based on the expression~\eqref{eq:Idelta}, which has 4 terms for each $\delta$ value. These four terms correspond to the four possible combinations for two HWPs with two angles each. Table~\ref{angle_prop1} shows the angles of these HWPs for each $\delta$ value used in our experiments.

\begin{table}[hptb]
	\begin{tabular}{|l|lllll|}
		\hline
		$\delta$ & $0.3$ & $0.4$ & $0.45$ & $0.5$ & $0.52$  \\ \hline
		$HWP_{A1}$ & $0$ & $0$ & $0$ & $0$ & $0$      \\
		$HWP_{A2}$ & $-63.20$ & $-61.77$ & $-61.05$ & $-60.34$ & $-60.05$ \\
		$HWP_{B1}$ & $22.5$ & $22.5$ & $22.5$ & $22.5$ & $22.5$       \\ 
        $HWP_{B2}$ & $85.70$ & $84.27$ & $83.55$ & $82.84$ & $82.55$  \\ \hline
	\end{tabular}
	\caption{HWP's angle for the expression~\eqref{eq:Idelta} for different values of $\delta$}
	\label{angle_prop1}
\end{table}

The second experiment is based on expression~\eqref{eq:Jgamma}. As with the previous expression, four combinations of two HWPs are required for each $\gamma$ value. Table~\ref{angle_prop2} groups these angles for each gamma used. 

\begin{table}[hptb]
	\begin{tabular}{|l|lll|}
		\hline
		$\gamma$ & $0$ & $\frac{\pi}{24}$ & $\frac{\pi}{12}$ \\ \hline
		$HWP_{A1}$ & $0$ & $0$ & $0$      \\
		$HWP_{A2}$ & $30$ & $26.25$ & $22.5$ \\
		$HWP_{B1}$ & $22.5$ & $16.88$ & $11.25$      \\ 
        $HWP_{B2}$ & $82.5$ & $80.63$ & $78.75$ \\ \hline
	\end{tabular}
	\caption{HWP's angle for the expression~\eqref{eq:Jgamma} for different values of $\delta$}
	\label{angle_prop2}
\end{table}

For these two equations, several values of each angle were possible, we have chosen to present only those used.

The value of the angles has been rounded to two digits, as we used computer-controlled high-precision motorized rotation stages to set the orientation of wave-plates with repeatability precision $0.02^{\circ}$

\begin{thebibliography}{30}%
	\makeatletter
	\providecommand \@ifxundefined [1]{%
		\@ifx{#1\undefined}
	}%
	\providecommand \@ifnum [1]{%
		\ifnum #1\expandafter \@firstoftwo
		\else \expandafter \@secondoftwo
		\fi
	}%
	\providecommand \@ifx [1]{%
		\ifx #1\expandafter \@firstoftwo
		\else \expandafter \@secondoftwo
		\fi
	}%
	\providecommand \natexlab [1]{#1}%
	\providecommand \enquote  [1]{``#1''}%
	\providecommand \bibnamefont  [1]{#1}%
	\providecommand \bibfnamefont [1]{#1}%
	\providecommand \citenamefont [1]{#1}%
	\providecommand \href@noop [0]{\@secondoftwo}%
	\providecommand \href [0]{\begingroup \@sanitize@url \@href}%
	\providecommand \@href[1]{\@@startlink{#1}\@@href}%
	\providecommand \@@href[1]{\endgroup#1\@@endlink}%
	\providecommand \@sanitize@url [0]{\catcode `\\12\catcode `\$12\catcode
		`\&12\catcode `\#12\catcode `\^12\catcode `\_12\catcode `\%12\relax}%
	\providecommand \@@startlink[1]{}%
	\providecommand \@@endlink[0]{}%
	\providecommand \url  [0]{\begingroup\@sanitize@url \@url }%
	\providecommand \@url [1]{\endgroup\@href {#1}{\urlprefix }}%
	\providecommand \urlprefix  [0]{URL }%
	\providecommand \Eprint [0]{\href }%
	\providecommand \doibase [0]{https://doi.org/}%
	\providecommand \selectlanguage [0]{\@gobble}%
	\providecommand \bibinfo  [0]{\@secondoftwo}%
	\providecommand \bibfield  [0]{\@secondoftwo}%
	\providecommand \translation [1]{[#1]}%
	\providecommand \BibitemOpen [0]{}%
	\providecommand \bibitemStop [0]{}%
	\providecommand \bibitemNoStop [0]{.\EOS\space}%
	\providecommand \EOS [0]{\spacefactor3000\relax}%
	\providecommand \BibitemShut  [1]{\csname bibitem#1\endcsname}%
	\let\auto@bib@innerbib\@empty
	\bibitem [{\citenamefont {Eastlake~3rd}\ \emph {et~al.}(2005)\citenamefont
		{Eastlake~3rd}, \citenamefont {Schiller},\ and\ \citenamefont
		{Crocker}}]{rfc4086}%
	\BibitemOpen
	\bibfield  {author} {\bibinfo {author} {\bibfnamefont {D.}~\bibnamefont
			{Eastlake~3rd}}, \bibinfo {author} {\bibfnamefont {J.}~\bibnamefont
			{Schiller}},\ and\ \bibinfo {author} {\bibfnamefont {S.}~\bibnamefont
			{Crocker}},\ }\href@noop {} {\emph {\bibinfo {title} {Randomness requirements
				for security}}},\ \bibinfo {type} {Tech. Rep.}\ (\bibinfo {year}
	{2005})\BibitemShut {NoStop}%
	\bibitem [{\citenamefont {Yuan}\ \emph {et~al.}(2015)\citenamefont {Yuan},
		\citenamefont {Zhou}, \citenamefont {Cao},\ and\ \citenamefont
		{Ma}}]{yuan2015intrinsic}%
	\BibitemOpen
	\bibfield  {author} {\bibinfo {author} {\bibfnamefont {X.}~\bibnamefont
			{Yuan}}, \bibinfo {author} {\bibfnamefont {H.}~\bibnamefont {Zhou}}, \bibinfo
		{author} {\bibfnamefont {Z.}~\bibnamefont {Cao}},\ and\ \bibinfo {author}
		{\bibfnamefont {X.}~\bibnamefont {Ma}},\ }\bibfield  {title} {\bibinfo
		{title} {Intrinsic randomness as a measure of quantum coherence},\
	}\href@noop {} {\bibfield  {journal} {\bibinfo  {journal} {Physical Review
				A}\ }\textbf {\bibinfo {volume} {92}},\ \bibinfo {pages} {022124} (\bibinfo
		{year} {2015})}\BibitemShut {NoStop}%
	\bibitem [{\citenamefont {Bell}(1964)}]{bell1964einstein}%
	\BibitemOpen
	\bibfield  {author} {\bibinfo {author} {\bibfnamefont {J.~S.}\ \bibnamefont
			{Bell}},\ }\bibfield  {title} {\bibinfo {title} {{On the Einstein Podolsky
				Rosen paradox}},\ }\href@noop {} {\bibfield  {journal} {\bibinfo  {journal}
			{Physics Physique Fizika}\ }\textbf {\bibinfo {volume} {1}},\ \bibinfo
		{pages} {195} (\bibinfo {year} {1964})}\BibitemShut {NoStop}%
	\bibitem [{\citenamefont {Mayers}\ and\ \citenamefont
		{Yao}(1998)}]{mayers1998quantum}%
	\BibitemOpen
	\bibfield  {author} {\bibinfo {author} {\bibfnamefont {D.}~\bibnamefont
			{Mayers}}\ and\ \bibinfo {author} {\bibfnamefont {A.}~\bibnamefont {Yao}},\
	}\bibfield  {title} {\bibinfo {title} {Quantum cryptography with imperfect
			apparatus},\ }in\ \href@noop {} {\emph {\bibinfo {booktitle} {Proceedings
				39th Annual Symposium on Foundations of Computer Science (Cat. No.
				98CB36280)}}}\ (\bibinfo {organization} {IEEE},\ \bibinfo {year} {1998})\
	pp.\ \bibinfo {pages} {503--509}\BibitemShut {NoStop}%
	\bibitem [{\citenamefont {Pironio}\ \emph
		{et~al.}(2010{\natexlab{a}})\citenamefont {Pironio}, \citenamefont
		{Ac{\'\i}n}, \citenamefont {Massar}, \citenamefont {de~La~Giroday},
		\citenamefont {Matsukevich}, \citenamefont {Maunz}, \citenamefont
		{Olmschenk}, \citenamefont {Hayes}, \citenamefont {Luo}, \citenamefont
		{Manning} \emph {et~al.}}]{pironio2010random}%
	\BibitemOpen
	\bibfield  {author} {\bibinfo {author} {\bibfnamefont {S.}~\bibnamefont
			{Pironio}}, \bibinfo {author} {\bibfnamefont {A.}~\bibnamefont {Ac{\'\i}n}},
		\bibinfo {author} {\bibfnamefont {S.}~\bibnamefont {Massar}}, \bibinfo
		{author} {\bibfnamefont {A.~B.}\ \bibnamefont {de~La~Giroday}}, \bibinfo
		{author} {\bibfnamefont {D.~N.}\ \bibnamefont {Matsukevich}}, \bibinfo
		{author} {\bibfnamefont {P.}~\bibnamefont {Maunz}}, \bibinfo {author}
		{\bibfnamefont {S.}~\bibnamefont {Olmschenk}}, \bibinfo {author}
		{\bibfnamefont {D.}~\bibnamefont {Hayes}}, \bibinfo {author} {\bibfnamefont
			{L.}~\bibnamefont {Luo}}, \bibinfo {author} {\bibfnamefont {T.~A.}\
			\bibnamefont {Manning}}, \emph {et~al.},\ }\bibfield  {title} {\bibinfo
		{title} {Random numbers certified by bell’s theorem},\ }\href@noop {}
	{\bibfield  {journal} {\bibinfo  {journal} {Nature}\ }\textbf {\bibinfo
			{volume} {464}},\ \bibinfo {pages} {1021} (\bibinfo {year}
		{2010}{\natexlab{a}})}\BibitemShut {NoStop}%
	\bibitem [{\citenamefont {Dupuis}\ \emph {et~al.}(2020)\citenamefont {Dupuis},
		\citenamefont {Fawzi},\ and\ \citenamefont {Renner}}]{dupuis2020entropy}%
	\BibitemOpen
	\bibfield  {author} {\bibinfo {author} {\bibfnamefont {F.}~\bibnamefont
			{Dupuis}}, \bibinfo {author} {\bibfnamefont {O.}~\bibnamefont {Fawzi}},\ and\
		\bibinfo {author} {\bibfnamefont {R.}~\bibnamefont {Renner}},\ }\bibfield
	{title} {\bibinfo {title} {Entropy accumulation},\ }\href@noop {} {\bibfield
		{journal} {\bibinfo  {journal} {Communications in Mathematical Physics}\
		}\textbf {\bibinfo {volume} {379}},\ \bibinfo {pages} {867} (\bibinfo {year}
		{2020})},\ \Eprint {https://arxiv.org/abs/arXiv:1607.01796}
	{arXiv:1607.01796} \BibitemShut {NoStop}%
	\bibitem [{\citenamefont {Arnon-Friedman}\ \emph {et~al.}(2019)\citenamefont
		{Arnon-Friedman}, \citenamefont {Renner},\ and\ \citenamefont
		{Vidick}}]{arnon2019simple}%
	\BibitemOpen
	\bibfield  {author} {\bibinfo {author} {\bibfnamefont {R.}~\bibnamefont
			{Arnon-Friedman}}, \bibinfo {author} {\bibfnamefont {R.}~\bibnamefont
			{Renner}},\ and\ \bibinfo {author} {\bibfnamefont {T.}~\bibnamefont
			{Vidick}},\ }\bibfield  {title} {\bibinfo {title} {Simple and tight
			device-independent security proofs},\ }\href@noop {} {\bibfield  {journal}
		{\bibinfo  {journal} {SIAM Journal on Computing}\ }\textbf {\bibinfo {volume}
			{48}},\ \bibinfo {pages} {181} (\bibinfo {year} {2019})}\BibitemShut
	{NoStop}%
	\bibitem [{\citenamefont {Arnon-Friedman}\ \emph {et~al.}(2018)\citenamefont
		{Arnon-Friedman}, \citenamefont {Dupuis}, \citenamefont {Fawzi},
		\citenamefont {Renner},\ and\ \citenamefont {Vidick}}]{arnon2018practical}%
	\BibitemOpen
	\bibfield  {author} {\bibinfo {author} {\bibfnamefont {R.}~\bibnamefont
			{Arnon-Friedman}}, \bibinfo {author} {\bibfnamefont {F.}~\bibnamefont
			{Dupuis}}, \bibinfo {author} {\bibfnamefont {O.}~\bibnamefont {Fawzi}},
		\bibinfo {author} {\bibfnamefont {R.}~\bibnamefont {Renner}},\ and\ \bibinfo
		{author} {\bibfnamefont {T.}~\bibnamefont {Vidick}},\ }\bibfield  {title}
	{\bibinfo {title} {Practical device-independent quantum cryptography via
			entropy accumulation},\ }\href@noop {} {\bibfield  {journal} {\bibinfo
			{journal} {Nature Communications}\ }\textbf {\bibinfo {volume} {9}},\
		\bibinfo {pages} {459} (\bibinfo {year} {2018})}\BibitemShut {NoStop}%
	\bibitem [{\citenamefont {Mironowicz}\ and\ \citenamefont
		{Paw{\l}owski}(2013)}]{mironowicz2013robustness}%
	\BibitemOpen
	\bibfield  {author} {\bibinfo {author} {\bibfnamefont {P.}~\bibnamefont
			{Mironowicz}}\ and\ \bibinfo {author} {\bibfnamefont {M.}~\bibnamefont
			{Paw{\l}owski}},\ }\bibfield  {title} {\bibinfo {title} {Robustness of
			quantum-randomness expansion protocols in the presence of noise},\
	}\href@noop {} {\bibfield  {journal} {\bibinfo  {journal} {Physical Review
				A}\ }\textbf {\bibinfo {volume} {88}},\ \bibinfo {pages} {032319} (\bibinfo
		{year} {2013})}\BibitemShut {NoStop}%
	\bibitem [{\citenamefont {Wooltorton}\ \emph {et~al.}(2022)\citenamefont
		{Wooltorton}, \citenamefont {Brown},\ and\ \citenamefont
		{Colbeck}}]{wooltorton2022tight}%
	\BibitemOpen
	\bibfield  {author} {\bibinfo {author} {\bibfnamefont {L.}~\bibnamefont
			{Wooltorton}}, \bibinfo {author} {\bibfnamefont {P.}~\bibnamefont {Brown}},\
		and\ \bibinfo {author} {\bibfnamefont {R.}~\bibnamefont {Colbeck}},\
	}\bibfield  {title} {\bibinfo {title} {Tight analytic bound on the trade-off
			between device-independent randomness and nonlocality},\ }\href@noop {}
	{\bibfield  {journal} {\bibinfo  {journal} {Physical Review Letters}\
		}\textbf {\bibinfo {volume} {129}},\ \bibinfo {pages} {150403} (\bibinfo
		{year} {2022})}\BibitemShut {NoStop}%
	\bibitem [{\citenamefont {Brown}\ \emph {et~al.}(2019)\citenamefont {Brown},
		\citenamefont {Ragy},\ and\ \citenamefont {Colbeck}}]{brown2019framework}%
	\BibitemOpen
	\bibfield  {author} {\bibinfo {author} {\bibfnamefont {P.~J.}\ \bibnamefont
			{Brown}}, \bibinfo {author} {\bibfnamefont {S.}~\bibnamefont {Ragy}},\ and\
		\bibinfo {author} {\bibfnamefont {R.}~\bibnamefont {Colbeck}},\ }\bibfield
	{title} {\bibinfo {title} {A framework for quantum-secure device-independent
			randomness expansion},\ }\href@noop {} {\bibfield  {journal} {\bibinfo
			{journal} {IEEE Transactions on Information Theory}\ }\textbf {\bibinfo
			{volume} {66}},\ \bibinfo {pages} {2964} (\bibinfo {year}
		{2019})}\BibitemShut {NoStop}%
	\bibitem [{\citenamefont {Brown}(2019)}]{brown2019constructions}%
	\BibitemOpen
	\bibfield  {author} {\bibinfo {author} {\bibfnamefont {P.}~\bibnamefont
			{Brown}},\ }\emph {\bibinfo {title} {On constructions of quantum-secure
			device-independent randomness expansion protocols}},\ \href@noop {} {Ph.D.
		thesis},\ \bibinfo  {school} {University of York} (\bibinfo {year}
	{2019})\BibitemShut {NoStop}%
	\bibitem [{\citenamefont {Vandenberghe}\ and\ \citenamefont
		{Boyd}(1996)}]{vandenberghe1996semidefinite}%
	\BibitemOpen
	\bibfield  {author} {\bibinfo {author} {\bibfnamefont {L.}~\bibnamefont
			{Vandenberghe}}\ and\ \bibinfo {author} {\bibfnamefont {S.}~\bibnamefont
			{Boyd}},\ }\bibfield  {title} {\bibinfo {title} {Semidefinite programming},\
	}\href@noop {} {\bibfield  {journal} {\bibinfo  {journal} {SIAM review}\
		}\textbf {\bibinfo {volume} {38}},\ \bibinfo {pages} {49} (\bibinfo {year}
		{1996})}\BibitemShut {NoStop}%
	\bibitem [{\citenamefont {Navascu{\'e}s}\ \emph {et~al.}(2007)\citenamefont
		{Navascu{\'e}s}, \citenamefont {Pironio},\ and\ \citenamefont
		{Ac{\'\i}n}}]{navascues2007bounding}%
	\BibitemOpen
	\bibfield  {author} {\bibinfo {author} {\bibfnamefont {M.}~\bibnamefont
			{Navascu{\'e}s}}, \bibinfo {author} {\bibfnamefont {S.}~\bibnamefont
			{Pironio}},\ and\ \bibinfo {author} {\bibfnamefont {A.}~\bibnamefont
			{Ac{\'\i}n}},\ }\bibfield  {title} {\bibinfo {title} {Bounding the set of
			quantum correlations},\ }\href@noop {} {\bibfield  {journal} {\bibinfo
			{journal} {Physical Review Letters}\ }\textbf {\bibinfo {volume} {98}},\
		\bibinfo {pages} {010401} (\bibinfo {year} {2007})}\BibitemShut {NoStop}%
	\bibitem [{\citenamefont {Navascu{\'e}s}\ \emph {et~al.}(2008)\citenamefont
		{Navascu{\'e}s}, \citenamefont {Pironio},\ and\ \citenamefont
		{Ac{\'\i}n}}]{navascues2008convergent}%
	\BibitemOpen
	\bibfield  {author} {\bibinfo {author} {\bibfnamefont {M.}~\bibnamefont
			{Navascu{\'e}s}}, \bibinfo {author} {\bibfnamefont {S.}~\bibnamefont
			{Pironio}},\ and\ \bibinfo {author} {\bibfnamefont {A.}~\bibnamefont
			{Ac{\'\i}n}},\ }\bibfield  {title} {\bibinfo {title} {A convergent hierarchy
			of semidefinite programs characterizing the set of quantum correlations},\
	}\href@noop {} {\bibfield  {journal} {\bibinfo  {journal} {New Journal of
				Physics}\ }\textbf {\bibinfo {volume} {10}},\ \bibinfo {pages} {073013}
		(\bibinfo {year} {2008})}\BibitemShut {NoStop}%
	\bibitem [{\citenamefont {Impagliazzo}\ \emph {et~al.}(1989)\citenamefont
		{Impagliazzo}, \citenamefont {Levin},\ and\ \citenamefont
		{Luby}}]{impagliazzo1989pseudo}%
	\BibitemOpen
	\bibfield  {author} {\bibinfo {author} {\bibfnamefont {R.}~\bibnamefont
			{Impagliazzo}}, \bibinfo {author} {\bibfnamefont {L.~A.}\ \bibnamefont
			{Levin}},\ and\ \bibinfo {author} {\bibfnamefont {M.}~\bibnamefont {Luby}},\
	}\bibfield  {title} {\bibinfo {title} {Pseudo-random generation from one-way
			functions},\ }in\ \href@noop {} {\emph {\bibinfo {booktitle} {Proceedings of
				the twenty-first annual ACM symposium on Theory of computing}}}\ (\bibinfo
	{year} {1989})\ pp.\ \bibinfo {pages} {12--24}\BibitemShut {NoStop}%
	\bibitem [{\citenamefont {Konig}\ \emph {et~al.}(2009)\citenamefont {Konig},
		\citenamefont {Renner},\ and\ \citenamefont
		{Schaffner}}]{konig2009operational}%
	\BibitemOpen
	\bibfield  {author} {\bibinfo {author} {\bibfnamefont {R.}~\bibnamefont
			{Konig}}, \bibinfo {author} {\bibfnamefont {R.}~\bibnamefont {Renner}},\ and\
		\bibinfo {author} {\bibfnamefont {C.}~\bibnamefont {Schaffner}},\ }\bibfield
	{title} {\bibinfo {title} {The operational meaning of min-and max-entropy},\
	}\href@noop {} {\bibfield  {journal} {\bibinfo  {journal} {IEEE Transactions
				on Information theory}\ }\textbf {\bibinfo {volume} {55}},\ \bibinfo {pages}
		{4337} (\bibinfo {year} {2009})}\BibitemShut {NoStop}%
	\bibitem [{\citenamefont {Clauser}\ \emph {et~al.}(1969)\citenamefont
		{Clauser}, \citenamefont {Horne}, \citenamefont {Shimony},\ and\
		\citenamefont {Holt}}]{clauser1969proposed}%
	\BibitemOpen
	\bibfield  {author} {\bibinfo {author} {\bibfnamefont {J.~F.}\ \bibnamefont
			{Clauser}}, \bibinfo {author} {\bibfnamefont {M.~A.}\ \bibnamefont {Horne}},
		\bibinfo {author} {\bibfnamefont {A.}~\bibnamefont {Shimony}},\ and\ \bibinfo
		{author} {\bibfnamefont {R.~A.}\ \bibnamefont {Holt}},\ }\bibfield  {title}
	{\bibinfo {title} {Proposed experiment to test local hidden-variable
			theories},\ }\href@noop {} {\bibfield  {journal} {\bibinfo  {journal}
			{Physical review letters}\ }\textbf {\bibinfo {volume} {23}},\ \bibinfo
		{pages} {880} (\bibinfo {year} {1969})}\BibitemShut {NoStop}%
	\bibitem [{\citenamefont {{\v{S}}upi{\'c}}\ and\ \citenamefont
		{Bowles}(2020)}]{vsupic2020self}%
	\BibitemOpen
	\bibfield  {author} {\bibinfo {author} {\bibfnamefont {I.}~\bibnamefont
			{{\v{S}}upi{\'c}}}\ and\ \bibinfo {author} {\bibfnamefont {J.}~\bibnamefont
			{Bowles}},\ }\bibfield  {title} {\bibinfo {title} {Self-testing of quantum
			systems: a review},\ }\href@noop {} {\bibfield  {journal} {\bibinfo
			{journal} {Quantum}\ }\textbf {\bibinfo {volume} {4}},\ \bibinfo {pages}
		{337} (\bibinfo {year} {2020})}\BibitemShut {NoStop}%
	\bibitem [{\citenamefont {Miller}\ and\ \citenamefont
		{Shi}(2017)}]{miller2017universal}%
	\BibitemOpen
	\bibfield  {author} {\bibinfo {author} {\bibfnamefont {C.~A.}\ \bibnamefont
			{Miller}}\ and\ \bibinfo {author} {\bibfnamefont {Y.}~\bibnamefont {Shi}},\
	}\bibfield  {title} {\bibinfo {title} {Universal security for randomness
			expansion from the spot-checking protocol},\ }\href@noop {} {\bibfield
		{journal} {\bibinfo  {journal} {SIAM Journal on Computing}\ }\textbf
		{\bibinfo {volume} {46}},\ \bibinfo {pages} {1304} (\bibinfo {year}
		{2017})}\BibitemShut {NoStop}%
	\bibitem [{\citenamefont {Fawzi}\ \emph {et~al.}(2019)\citenamefont {Fawzi},
		\citenamefont {Saunderson},\ and\ \citenamefont
		{Parrilo}}]{fawzi2019semidefinite}%
	\BibitemOpen
	\bibfield  {author} {\bibinfo {author} {\bibfnamefont {H.}~\bibnamefont
			{Fawzi}}, \bibinfo {author} {\bibfnamefont {J.}~\bibnamefont {Saunderson}},\
		and\ \bibinfo {author} {\bibfnamefont {P.~A.}\ \bibnamefont {Parrilo}},\
	}\bibfield  {title} {\bibinfo {title} {Semidefinite approximations of the
			matrix logarithm},\ }\href@noop {} {\bibfield  {journal} {\bibinfo  {journal}
			{Foundations of Computational Mathematics}\ }\textbf {\bibinfo {volume}
			{19}},\ \bibinfo {pages} {259} (\bibinfo {year} {2019})}\BibitemShut
	{NoStop}%
	\bibitem [{\citenamefont {Fawzi}\ and\ \citenamefont
		{Fawzi}(2018)}]{fawzi2018efficient}%
	\BibitemOpen
	\bibfield  {author} {\bibinfo {author} {\bibfnamefont {H.}~\bibnamefont
			{Fawzi}}\ and\ \bibinfo {author} {\bibfnamefont {O.}~\bibnamefont {Fawzi}},\
	}\bibfield  {title} {\bibinfo {title} {Efficient optimization of the quantum
			relative entropy},\ }\href@noop {} {\bibfield  {journal} {\bibinfo  {journal}
			{Journal of Physics A: Mathematical and Theoretical}\ }\textbf {\bibinfo
			{volume} {51}},\ \bibinfo {pages} {154003} (\bibinfo {year}
		{2018})}\BibitemShut {NoStop}%
	\bibitem [{\citenamefont {Brown}\ \emph {et~al.}(2021)\citenamefont {Brown},
		\citenamefont {Fawzi},\ and\ \citenamefont {Fawzi}}]{brown2021device}%
	\BibitemOpen
	\bibfield  {author} {\bibinfo {author} {\bibfnamefont {P.}~\bibnamefont
			{Brown}}, \bibinfo {author} {\bibfnamefont {H.}~\bibnamefont {Fawzi}},\ and\
		\bibinfo {author} {\bibfnamefont {O.}~\bibnamefont {Fawzi}},\ }\bibfield
	{title} {\bibinfo {title} {{Device-independent lower bounds on the
				conditional von Neumann entropy}},\ }\href@noop {} {\bibfield  {journal}
		{\bibinfo  {journal} {arXiv preprint arXiv:2106.13692}\ } (\bibinfo {year}
		{2021})}\BibitemShut {NoStop}%
	\bibitem [{\citenamefont {Pironio}\ \emph
		{et~al.}(2010{\natexlab{b}})\citenamefont {Pironio}, \citenamefont
		{Navascu{\'e}s},\ and\ \citenamefont {Acin}}]{pironio2010convergent}%
	\BibitemOpen
	\bibfield  {author} {\bibinfo {author} {\bibfnamefont {S.}~\bibnamefont
			{Pironio}}, \bibinfo {author} {\bibfnamefont {M.}~\bibnamefont
			{Navascu{\'e}s}},\ and\ \bibinfo {author} {\bibfnamefont {A.}~\bibnamefont
			{Acin}},\ }\bibfield  {title} {\bibinfo {title} {Convergent relaxations of
			polynomial optimization problems with noncommuting variables},\ }\href@noop
	{} {\bibfield  {journal} {\bibinfo  {journal} {SIAM Journal on Optimization}\
		}\textbf {\bibinfo {volume} {20}},\ \bibinfo {pages} {2157} (\bibinfo {year}
		{2010}{\natexlab{b}})}\BibitemShut {NoStop}%
	\bibitem [{\citenamefont {Wittek}(2015)}]{wittek2015algorithm}%
	\BibitemOpen
	\bibfield  {author} {\bibinfo {author} {\bibfnamefont {P.}~\bibnamefont
			{Wittek}},\ }\bibfield  {title} {\bibinfo {title} {Algorithm 950:
			Ncpol2sdpa—sparse semidefinite programming relaxations for polynomial
			optimization problems of noncommuting variables},\ }\href@noop {} {\bibfield
		{journal} {\bibinfo  {journal} {ACM Transactions on Mathematical Software
				(TOMS)}\ }\textbf {\bibinfo {volume} {41}},\ \bibinfo {pages} {1} (\bibinfo
		{year} {2015})}\BibitemShut {NoStop}%
	\bibitem [{\citenamefont {Brown}(2022)}]{Brown2022}%
	\BibitemOpen
	\bibfield  {author} {\bibinfo {author} {\bibfnamefont {P.~J.}\ \bibnamefont
			{Brown}},\ }\href@noop {} {\bibinfo {title} {Example scripts for computing
			rates of device-independent protocols}},\ \bibinfo {howpublished}
	{\url{https://github.com/peterjbrown519/DI-rates}} (\bibinfo {year}
	{2022})\BibitemShut {NoStop}%
	\bibitem [{\citenamefont {Nieto-Silleras}\ \emph {et~al.}(2014)\citenamefont
		{Nieto-Silleras}, \citenamefont {Pironio},\ and\ \citenamefont
		{Silman}}]{nieto2014using}%
	\BibitemOpen
	\bibfield  {author} {\bibinfo {author} {\bibfnamefont {O.}~\bibnamefont
			{Nieto-Silleras}}, \bibinfo {author} {\bibfnamefont {S.}~\bibnamefont
			{Pironio}},\ and\ \bibinfo {author} {\bibfnamefont {J.}~\bibnamefont
			{Silman}},\ }\bibfield  {title} {\bibinfo {title} {Using complete measurement
			statistics for optimal device-independent randomness evaluation},\
	}\href@noop {} {\bibfield  {journal} {\bibinfo  {journal} {New Journal of
				Physics}\ }\textbf {\bibinfo {volume} {16}},\ \bibinfo {pages} {013035}
		(\bibinfo {year} {2014})}\BibitemShut {NoStop}%
	\bibitem [{\citenamefont {Bancal}\ \emph {et~al.}(2014)\citenamefont {Bancal},
		\citenamefont {Sheridan},\ and\ \citenamefont {Scarani}}]{bancal2014more}%
	\BibitemOpen
	\bibfield  {author} {\bibinfo {author} {\bibfnamefont {J.-D.}\ \bibnamefont
			{Bancal}}, \bibinfo {author} {\bibfnamefont {L.}~\bibnamefont {Sheridan}},\
		and\ \bibinfo {author} {\bibfnamefont {V.}~\bibnamefont {Scarani}},\
	}\bibfield  {title} {\bibinfo {title} {More randomness from the same data},\
	}\href@noop {} {\bibfield  {journal} {\bibinfo  {journal} {New Journal of
				Physics}\ }\textbf {\bibinfo {volume} {16}},\ \bibinfo {pages} {033011}
		(\bibinfo {year} {2014})}\BibitemShut {NoStop}%
	\bibitem [{\citenamefont {Nieto-Silleras}\ \emph {et~al.}(2018)\citenamefont
		{Nieto-Silleras}, \citenamefont {Bamps}, \citenamefont {Silman},\ and\
		\citenamefont {Pironio}}]{nieto2018device}%
	\BibitemOpen
	\bibfield  {author} {\bibinfo {author} {\bibfnamefont {O.}~\bibnamefont
			{Nieto-Silleras}}, \bibinfo {author} {\bibfnamefont {C.}~\bibnamefont
			{Bamps}}, \bibinfo {author} {\bibfnamefont {J.}~\bibnamefont {Silman}},\ and\
		\bibinfo {author} {\bibfnamefont {S.}~\bibnamefont {Pironio}},\ }\bibfield
	{title} {\bibinfo {title} {{Device-independent randomness generation from
				several Bell estimators}},\ }\href@noop {} {\bibfield  {journal} {\bibinfo
			{journal} {New Journal of Physics}\ }\textbf {\bibinfo {volume} {20}},\
		\bibinfo {pages} {023049} (\bibinfo {year} {2018})}\BibitemShut {NoStop}%
	\bibitem [{\citenamefont {ApS}(2021)}]{mosek21}%
	\BibitemOpen
	\bibfield  {author} {\bibinfo {author} {\bibfnamefont {M.}~\bibnamefont
			{ApS}},\ }\href {https://docs.mosek.com/modeling-cookbook/index.html} {\emph
		{\bibinfo {title} {MOSEK Modeling Cookbook}}},\ \bibinfo {edition} {3rd}\
	ed.\ (\bibinfo {year} {2021})\BibitemShut {NoStop}%
\end{thebibliography}
\end{document}